\documentclass[conference]{IEEEtran}
\IEEEoverridecommandlockouts

\usepackage{cite}
\usepackage{amsmath,amssymb,amsfonts}
\usepackage{booktabs}
\usepackage{graphicx}
\usepackage{textcomp}
\usepackage{xcolor}
\usepackage{url}
\usepackage{pgfplots}
\pgfplotsset{compat=1.18}
\usetikzlibrary{arrows.meta,positioning}

\def\BibTeX{{\rm B\kern-.05em{\sc i\kern-.025em b}\kern-.08em
T\kern-.1667em\lower.7ex\hbox{E}\kern-.125emX}}

\begin{document}

\title{When Post-Anchor Metrics Fail: Stabilization Regimes in AI-Evidence Open-Source Projects}

\author{
\IEEEauthorblockN{Hudson Craig\textsuperscript{1},
Benjamin Barlog\textsuperscript{1}, Chenggang Wang\textsuperscript{2}, Longwei Wang\textsuperscript{3}, Zedong Peng\textsuperscript{1,\,*}}
\IEEEauthorblockA{%
\textsuperscript{1}Department of Computer Science, University of Montana, Missoula, MT, USA\\
\textsuperscript{2}Polytechnic Institute, University of Oklahoma, Tulsa, OK, USA\\
\textsuperscript{3}Department of Computer Science, University of South Dakota, Vermillion, SD, USA\\[2pt]
hudson.craig@umconnect.umt.edu, benjamin.barlog@umconnect.umt.edu\\
chenggang.wang@ou.edu, longwei.wang@usd.edu, zedong.peng@umt.edu\\[2pt]
\textsuperscript{*}Corresponding author}
}

\maketitle

\begin{abstract}
Repository mining studies increasingly analyze AI-evidence projects, yet it remains unclear how to measure whether architectural changes create deferred stabilization obligations. A natural metric, post-anchor stabilization density, counts tests, CI gates, documentation, and fixes appearing after a durable boundary is introduced. We show that this metric fails. In a diff-level study of 338 high-visibility 2026 GitHub repositories, anchors are common (321 of 338 contain real changed-file anchor evidence), but a controlled 308-event contiguous-window experiment finds no post-anchor uplift: broad and strict stabilization signals both yield median post/pre density ratios near 1.0, and non-anchor controls are equally dense. We introduce \emph{stabilization regimes}, six recurring patterns that explain why the density metric fails, and use human validation to calibrate them. Two independent coders label 100 stratified candidate-anchor events from blinded packets ($\kappa = 0.50$ on debt attribution). The validation exposes a two-layer trap: many candidate anchors are not durable boundaries (45 of 100), and even among valid anchors in this calibration sample the no-uplift result holds: only 3 of 100 events survive as attributable delayed obligations; the remaining 97 are explained by classifier error, anchor-local hardening, background maintenance, or pre-anchor hardening. Post-anchor density conflates pervasive maintenance with genuine debt; controlled designs with regime-aware attribution are necessary before repository mining can reliably identify stabilization obligations.
\end{abstract}


\section{Introduction}

A developer asks an AI coding agent for a bounded change: add support for a new model provider. The patch lands quickly, touching a provider adapter, a runtime configuration file, a workflow step, and a documentation note. Over the next few commits the repository gains a test fixture, a CI adjustment, a fallback guard, and a dependency pin. From a distance, this looks like deferred technical debt~\cite{cunningham1992wycash,kruchten2012metaphor}: an architectural boundary appears, and stabilization work follows.

But the inference may be wrong. The repository may already be in a hardening period. Non-anchor commits may show the same density of tests and CI changes~\cite{herzig2013tangled}. Or the stabilizers may have arrived with the boundary change itself~\cite{wen2022quickremedy}. A post-anchor count alone cannot distinguish these cases, and the distinction matters: a tool or reviewer that treats every dense post-anchor neighborhood as evidence of debt may trigger unnecessary repairs or scope-expanding changes the developer never requested~\cite{bacchelli2013code,rigby2013convergent,tufano2021codereview,pornprasit2024codereview}.

We study this ambiguity in 338 high-visibility AI-evidence repositories~\cite{kalliamvakou2014github,munaiah2017curating} collected in 2026. We call a durable structural or coordination boundary an \emph{anchor}: provider interfaces, workflow engines, schemas, runtime contracts, or agent-facing project instructions~\cite{verdecchia2020atd,sculley2015mltd}. The first question was whether AI-era repositories lack such boundaries. They do not: real changed-file diffs show that 321 of 338 repositories (95.0\%) contain anchor evidence. The problem is not missing anchors but what happens after them.

That question motivates the construct we study: \emph{post-anchor stabilization debt}, the deferred hardening obligation that becomes visible when a boundary is introduced without the tests, gates, guards, and documentation it needs. A controlled 308-event contiguous-window experiment finds no detectable uplift: broad and strict stabilization signals both produce median post/pre density ratios near~1.0, and non-anchor control windows are equally dense~\cite{lehman1980laws}. The null does not mean debt never occurs; it means that a density threshold cannot separate a few genuine cases from pervasive background maintenance.

These findings define a measurement trap with two layers. First, automated anchor extraction misclassifies many candidate events (human validation finds that 45 of 100 candidates are not durable anchors). Second, even among valid anchors, post-anchor density does not exceed control density.

The paper asks a single question throughout: \emph{can post-anchor stabilization density measure delayed debt?} We falsify the metric, diagnose why it fails through six recurring failure modes, and quantify the overcall with human consensus labels. The result is three contributions:
\begin{itemize}
    \item \textbf{Measurement trap (falsification).} Post-anchor stabilization
    density is unreliable at two points: anchor extraction and density-based
    attribution. In a 308-event experiment, post-anchor density does not
    exceed pre-anchor or control density under broad, strict, or
    window-size checks.
    \item \textbf{Stabilization regimes (diagnosis).} A six-regime taxonomy
    (saturation, background maintenance, mixed, pre-anchor hardening,
    anchor-local hardening, and post-uplift) explains why density patterns
    are insufficient and identifies which failure mode a candidate event
    exhibits before semantic attribution. The regimes also motivate a
    review-time anchor-ledger workflow that we sketch, but do not evaluate,
    in Section~\ref{sec:discussion}.
    \item \textbf{Human-calibrated consensus labels (quantification).} Two
    independent coders label 100 stratified events from blinded packets
    ($\kappa = 0.50$ on debt attribution, $\kappa = 0.39$ on anchor
    validity). Only 3 of 100 survive as attributable delayed obligations;
    the remaining 97 have definitive non-debt explanations.
\end{itemize}

\begin{figure}[t]
\centering
\begin{tikzpicture}[
  every node/.style={font=\scriptsize},
  stage/.style={draw, rounded corners, fill=blue!5, align=left,
                text width=0.58\columnwidth, inner sep=3.5pt},
  arr/.style={-{Stealth[length=5pt]}, semithick},
  cnt/.style={font=\scriptsize\bfseries, text=blue!45!black, align=left}]
\node[stage] (s1) {\textbf{1. Sampling.} 2026 AI-evidence repositories;
  runnable surface, star and commit thresholds.};
\node[stage, below=4.5mm of s1] (s2) {\textbf{2. Diff validation.} Real
  changed files, not commit messages, decide anchor presence.};
\node[stage, below=4.5mm of s2] (s3) {\textbf{3. Event windows.} Anchor, pre,
  post, and two non-anchor control windows per event.};
\node[stage, below=4.5mm of s3] (s4) {\textbf{4. Regime diagnosis.} Six
  stabilization regimes flag each event's failure mode.};
\node[stage, below=4.5mm of s4] (s5) {\textbf{5. Human validation.} Two coders
  label a stratified sample; consensus resolves disagreements.};
\draw[arr] (s1) -- (s2) node[cnt, midway, right=1.5mm] {338 projects};
\draw[arr] (s2) -- (s3) node[cnt, midway, right=1.5mm] {321 anchored};
\draw[arr] (s3) -- (s4) node[cnt, midway, right=1.5mm] {308 events};
\draw[arr] (s4) -- (s5) node[cnt, midway, right=1.5mm] {100 coded\\$\rightarrow$ 3 debt};
\end{tikzpicture}
\caption{Layered measurement workflow. Each layer narrows the claim before the
next is allowed to call later stabilization work debt; the right-hand counts
show the funnel from 338 sampled projects to 3 attributable delayed
obligations.}
\label{fig:workflow}
\end{figure}
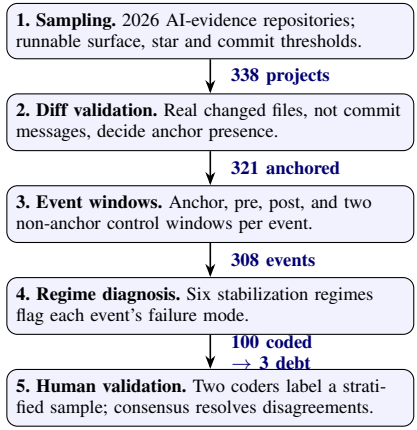

\section{Background and Related Work}

\subsection{Technical Debt and Stabilization}

Technical debt describes future maintenance cost incurred by short-term design or implementation tradeoffs~\cite{cunningham1992wycash,li2015technical,kruchten2012metaphor}; managing, prioritizing, and repaying it is a recurring challenge in long-lived projects~\cite{ernst2015measure,avgeriou2016dagstuhl,lenarduzzi2021prioritization,digkas2018payback}. Software-evolution theory cautions against a simple after-anchor story: Lehman's laws and later studies of software aging frame stabilization as part of the ordinary evolution loop rather than an exceptional cleanup phase~\cite{lehman1980laws,godfrey2000evolution}. Our event-study design therefore treats post-anchor work as a hypothesis to test: if pre-anchor and control windows are equally dense, a later hardening commit cannot automatically be called debt.

Release-engineering and continuous-integration research shows that build, test, and release automation are first-class quality mechanisms, not peripheral metadata~\cite{vasilescu2015ci,hilton2016ci,beller2017travis,zhao2017ci}. We therefore treat CI files, fixtures, release scripts, and workflow changes as stabilization evidence, but compare anchor windows against controls so that ordinary delivery activity is not mistaken for an anchor-induced obligation. Debt is also structural: hidden debt in machine-learning systems and architectural debt arise from interfaces, dependencies, and design decisions rather than local code smells alone~\cite{sculley2015mltd,verdecchia2020atd,tufano2015smell}. Our anchor construct captures such boundary events without a full architecture model. Its unit of analysis is neither a release phase, a comment, nor a code smell, but an anchor event and the hardening work that should have accompanied it.

\subsection{Self-Admitted and AI-Era Debt}

Self-admitted technical debt (SATD) is the closest methodological neighbor: text- and comment-mining studies detect developer-admitted debt and track how it is introduced, classified, removed, and repaid~\cite{potdar2014satd,bavota2016satd,huang2018satd,ren2019satd,sierra2019survey,alves2016tdmapping,maldonado2017removal,zampetti2017recommend}. Two lessons shape our design: debt labels are construct-heavy and need semantic judgment over local evidence (ours uses structured packets, a fixed schema, and a decision tree), and debt is temporal, so what matters is how an obligation appears, persists, and is repaid, not merely whether a debt-like phrase is present. Unlike SATD, our construct requires no admitting comment; it observes whether a durable boundary is introduced without its stabilizers, which suits AI-evidence repositories where debt surfaces in tests, workflows, configuration, and documentation.

Large language models are reshaping the development lifecycle~\cite{fan2023llm4se,russo2024genai,hou2024llm4se,wang2024testing}, and empirical work reports both productivity gains and new risks: changed search-and-validate behavior, added review burden, variable correctness, and insecure suggestions~\cite{vaithilingam2022copilot,ziegler2024copilot,nguyen2022copilot,dakhel2023copilot,liu2023evalplus,du2024classeval,pearce2022asleep}. Most of this work evaluates individual tasks rather than repository evolution after AI-mediated changes land. Closer to debt, recent studies characterize GenAI-induced and agent-authored SATD in code comments and ML software~\cite{mujahid2026gist,obrien2022mlsatd}. We avoid claiming verified AI authorship and instead study \emph{AI-evidence repositories} (projects with explicit AI-development evidence, provider/runtime assets, agent-facing documentation, or prompt/model artifacts), a scope weaker for attribution but better suited to evolution questions where AI and human work mix.

\subsection{Bots, Agents, and Repository Mining}

Software-bot and DevBot research shows projects delegating maintenance, review, testing, and coordination to automation whose output maintainers must still judge~\cite{wessel2018bots,erlenhov2020devbots,bouzenia2025repairagent}. AI coding agents intensify this: agent instructions, prompt libraries, model configuration, and workflow guards become coordination anchors when future work must respect them, which is why we treat them as candidate anchors but still require diff and stabilizer evidence before any debt claim.

Mining GitHub is error-prone because visibility, metadata, and mixed-quality projects can mislead conclusions~\cite{kalliamvakou2014github,munaiah2017curating}. Two findings directly motivate our controls: commits often bundle tangled, weakly related changes~\cite{herzig2013tangled,silva2016refactor}, and developers frequently follow a commit with quick-remedy fixes for omitted content~\cite{wen2022quickremedy}, so a post-anchor window cannot be read causally without baselines. Our own pipeline confirms a further risk: commit-message scouting overestimated missing and delayed anchors, so changed-file diffs, not messages, are the source of truth. In this spirit our contribution is not a classifier that ``solves'' stabilization debt but a measurement artifact that shows where the post-anchor shortcut breaks and supplies a cleaner unit of analysis: anchor event, expected stabilizers, pre/post windows, controls, and regime-aware attribution.

\section{Definitions}

\textbf{Anchor.} An anchor is a durable architecture or coordination boundary. We distinguish five anchor types: API/provider/runtime boundaries, workflow/engine boundaries, storage/schema/data boundaries, coordination contracts, and AI-runtime boundaries. Tests, CI workflows, release scripts, lockfiles, and ordinary README edits are not anchors by default.

\textbf{Stabilization work.} Stabilization work is observable hardening or repair work: tests, CI/release gates, documentation contracts, runtime guards, dependency/toolchain repairs, provider integrations, and UX regression repairs.

\textbf{Stabilization debt.} Stabilization debt is the delayed hardening obligation attributable to an anchor. A later test or fix is not automatically debt; it becomes debt evidence only when it can be tied to a boundary that was introduced without the support needed to evolve safely.

\textbf{Stabilization burden.} Stabilization burden is the measured intensity of stabilization work in a window: stabilization commits, stabilization files, category breadth, and severe categories.

Table~\ref{tab:schema} defines the measurement schema. Panel~(a) lists the stabilization categories used by the classifier and the annotation schema; they deliberately mix source, test, configuration, documentation, and release artifacts, because an anchor's hardening obligation is rarely isolated in one file type. Panel~(b) is the expected-stabilizer checklist per anchor type: not a learned model but an engineering derivation from common boundary responsibilities (a provider boundary, for instance, needs tests, integration, guards, and docs). Neither panel implies that every such file is debt. Debt attribution requires a temporal and semantic tie to an anchor, which is why the event study, controls, and human-validated labels are necessary; a later stabilizer counts as debt only if human validation connects it to the anchor.

\begin{table}[htbp]
\caption{Measurement schema. (a)~Stabilization categories and evidence
examples; (b)~expected stabilizers per anchor type.}
\label{tab:schema}
\centering
\scriptsize
\textbf{(a) Stabilization categories}\\[1.5pt]
\begin{tabular}{ll}
\toprule
Category & Evidence examples \\
\midrule
Test stabilization & tests, fixtures, mocks, regression suites \\
CI/release stabilization & workflows, builds, packaging, release gates \\
Documentation stabilization & contracts, usage docs, migration notes \\
Runtime hardening & guards, retries, validation, error handling \\
Dependency/toolchain & lockfiles, pins, tool configs, migrations \\
Integration/provider & auth, config, provider fallback, adapters \\
UX/product regression & UI fixes, accessibility, layout regressions \\
\bottomrule
\end{tabular}\\[4pt]
\textbf{(b) Expected stabilizers per anchor type}\\[1.5pt]
\begin{tabular}{ll}
\toprule
Anchor type & Expected stabilizers \\
\midrule
API/provider & Tests; integration; guards; docs \\
Workflow/engine & Tests; CI/release; guards \\
Storage/schema & Tests; guards; dependency/toolchain \\
Coordination contract & Docs/contracts; tests \\
AI runtime & Tests; integration; guards; docs \\
\bottomrule
\end{tabular}
\end{table}

\section{Study Design}

\subsection{Repository Pool}

We collected a year-to-date 2026 GitHub pool (snapshot as of June 2026) across media, productivity, knowledge/search, developer tools, workflow automation, education, research, health/accessibility, and business domains. The broad accepted pool freezes all repositories that passed basic screening before later balancing or category caps: public GitHub repositories created or first released in early-to-mid 2026, at least 100 stars, estimated 50 or more commits, a runnable product/tool surface, and AI-evidence signals. AI-evidence signals are explicit markers of AI-native development or runtime: agent-facing documentation, explicit AI-development evidence in commits or configuration, provider or agent-runtime assets, AI-native product identity, and prompt or model artifacts. Domain queries seeded the search; each candidate was then enriched with these signals and screened, narrowing 695 enriched candidates to the 338-project pool (Table~\ref{tab:funnel}).

The search protocol deliberately did not restrict the pool to coding-agent projects. Coding agents and developer tools are present because they are important AI-era software, but the pool also includes media tools, productivity tools, knowledge and search systems, workflow products, education tools, business/legal/finance tools, and consumer-facing products. We are not studying how coding agents themselves are implemented, but public projects that show AI-development or AI-runtime evidence while delivering a runnable software surface.

Screening excluded prompt collections, awesome lists, course-only repositories, starter templates, and skill-only packages unless they also delivered a concrete product or workflow. Rather than deleting devtool-like records after the fact (which can hide category bias), the accepted pool records devtool-like and non-devtool-like counts so later analysis can stratify. This pool is therefore a high-visibility AI-evidence product pool, not a random sample of all AI-assisted programming: the star and commit thresholds select for projects with enough public attention for commit-window analysis. The screening rule was fixed before the validation pass, so that changing the pool after seeing the stabilization result could not tune the phenomenon.

\begin{table}[htbp]
\caption{Accepted-pool funnel and evidence summary.}
\label{tab:funnel}
\centering
\begin{tabular}{lr}
\toprule
Metric & Count \\
\midrule
Enriched candidate records & 695 \\
Rejected after basic screening & 357 \\
Full accepted pool & 338 \\
Devtool-like accepted records & 168 \\
Non-devtool-like accepted records & 170 \\
Agent-facing docs present & 196 \\
Explicit AI-development evidence & 269 \\
Provider or agent runtime evidence & 303 \\
\bottomrule
\end{tabular}
\end{table}

\subsection{Evidence Modes}

The study uses two evidence modes. The first is a sampled-phase diff corpus from engineer validation over all 338 repositories. The second is a checkpointed contiguous-window experiment. For each selected candidate anchor event, we locate the anchor SHA on the repository's default branch, fetch a contiguous commit window around it, and fetch GitHub commit-detail diffs for the anchor, pre-anchor, post-anchor, random non-anchor, and matched non-anchor windows.

\subsection{Event Extraction}

The sampled-phase experiment scans the engineer-validation diffs and extracts one candidate anchor event per repository. We intentionally separate these events from the older AnchorFinder warning labels. AnchorFinder was useful as a scout, but the paper's events are derived from changed-file and patch evidence under the mutually exclusive taxonomy in Section III. This separation is important because the early warning logic treated tests, docs, CI, and release work as signals that an anchor might be missing. In the study taxonomy, those files usually become stabilizers rather than anchors.

For each repository, we selected the earliest commit on the default branch that satisfied the anchor definition and touched at least one durable boundary file, excluding commits whose only anchor-like files were tests, CI, release scripts, lockfiles, or ordinary README edits. This deterministic rule avoids author-selected ``salient'' events and ensures that the choice is reproducible from the diff corpus alone.

For each selected event, the contiguous experiment stores the repository, SHA, commit date, message, anchor type, anchor files, anchor diff size, and project metadata. We then query the repository's default branch and locate the anchor SHA in the branch history. GitHub commit-list order is newest to oldest, so the post-anchor window is the set of newer commits immediately before the anchor in the list, while the pre-anchor window is the set of older commits immediately after it. We use a 10-commit window on each side.

The resulting event-study population covers 308 of the 338 accepted repositories (91.1\%). Each included repository contributes exactly one selected candidate anchor event, so the population is not clustered by repository. Thirty repositories have no qualifying candidate anchor event under our extraction rules. We phrase event-study claims over the 308 included repositories and use the full 338 for anchor-presence counts.

\subsection{Controls}

A post-anchor window alone cannot distinguish delayed hardening from ordinary local maintenance. We therefore compute two controls in the same repository neighborhood. The random control is a deterministic random non-anchor commit within a local range of 30 commits around the anchor, seeded by the anchor SHA for reproducibility. The matched control is the non-anchor commit in that range whose diff size is closest to the anchor commit's diff size. Neither control window overlaps with the anchor, pre-anchor, or post-anchor windows. For each control, we measure the same forward-looking window. This design is still observational, but it prevents the paper from treating a dense maintenance region as an anchor-specific effect without comparison.

\subsection{Metrics and Effect Sizes}

For each window we compute two density measures. Commit density is the fraction of commits in the window with any stabilization signal. File density is the fraction of changed files in the window classified as stabilization files. File density is the primary measure because a single large commit can touch many unrelated files, and the burden of stabilizing an anchor is often visible in the spread of changed files across tests, workflows, docs, configuration, and runtime code.

The paired post/pre ratio uses a smoothed file-density ratio: $(post\_stab\_files + 1)/(post\_total\_files + 1)$ divided by $(pre\_stab\_files + 1)/(pre\_total\_files + 1)$. Smoothing avoids unstable division when a small pre-window has no stabilization files. We report the median ratio rather than the mean because window sizes and diff sizes are heavy-tailed in young GitHub projects. Bootstrap confidence intervals summarize uncertainty in the median, and Cliff's delta compares post-anchor file density with pre-anchor, random-control, and matched-control windows without assuming a normal distribution.

These choices are conservative for the claim we originally expected. If post-anchor stabilization debt were a large and systematic delayed wave, it should appear despite medians, smoothing, strict signals, and controls. The fact that it does not appear under these checks is therefore informative: the phenomenon is not absent, but it is not captured by a simple aggregate uplift.

\subsection{Diagnostic Regimes}

The pattern diagnostic is a post-hoc interpretability layer, not a substitute for human validation. It assigns each event to one of six regimes using deterministic rules applied in priority order over the five window densities and the smoothed post/pre ratio. A window is classified as \emph{dense} when its broad file density is at least 0.70.
\begin{enumerate}
\item \emph{Pre-anchor stabilization}: the smoothed post/pre ratio is below
  0.75, indicating pre-anchor density dominates.
\item \emph{Anchor-local only}: the anchor commit is dense but the post-window
  density is below 0.50.
\item \emph{Post-anchor uplift}: post density exceeds both pre and both
  controls by at least 0.05 and the ratio exceeds 1.50.
\item \emph{Stabilization saturation}: anchor, pre, post, and both controls
  are all dense.
\item \emph{Background maintenance}: the post window is dense and at least one
  non-anchor control has comparable or higher density.
\item \emph{Mixed or unclear}: all remaining events, including those with
  missing controls.
\end{enumerate}
These labels are deliberately conservative. Their purpose is to route examples into schema-bound human review and prevent overclaiming.

The diagnostic layer is also where the paper's terminology becomes disciplined. Only post-anchor uplift even superficially resembles the simple delayed-debt story---and our human validation later shows that even this resemblance is misleading, since the genuine delayed obligations do not fall in that regime. Saturation and background maintenance are still important engineering states, but they should not be collapsed into delayed debt. Pre-anchor stabilization suggests anticipatory hardening. Anchor-local only suggests synchronous hardening. Mixed cases require patch-level review. This separation lets the study preserve the new term ``stabilization debt'' without using it as a blanket label for every later test or release commit.

\subsection{Broad and Strict Signals}

The broad signal is designed for recall. It counts common stabilization evidence such as tests, CI/release files, docs, guards, fallback/error handling, dependency/toolchain files, provider/auth/config integration, and UI regression terms. Because broad signals can overcount generic maintenance, we also run a strict sensitivity analysis. The strict signal excludes generic ``fix'', broad API terms, and broad UI terms unless the file path or patch contains stronger stabilization evidence such as tests, workflows, architecture contracts, guards, retries, lockfiles, auth/config integration, or concrete UI regression terms.

We report both signals because either one alone would be misleading. The broad signal is appropriate for queue construction: it avoids missing cases where stabilization is spread across docs, CI, provider glue, and UI surfaces. The strict signal is appropriate for skepticism: it asks whether the same interpretation survives after generic maintenance terms are removed. The answer is yes: absolute densities change, but the conclusion does not become a strong post-anchor uplift.

\subsection{Human Validation Design}

The automated pipeline produces diagnostic regimes, not final debt labels. To obtain consensus-based debt attribution and calibrate the regime diagnosis, we drew a stratified 100-case sample from the 308-event pool, preserving all rare regimes (post-uplift, anchor-local) and sampling common ones (saturation, background maintenance, mixed). For each case, the validation package stores the anchor URL, anchor files, commit message, pre/post/control densities, changed-file samples, and project metadata---but hides all automated diagnostic labels.

Two independent coders received blinded HTML packets presenting the 100 cases in different randomized orders. Each coder assigned anchor validity, anchor type, debt attribution, and evidence confidence following a decision tree that required inspecting the control comparison before assigning a delayed-obligation label, though counterfactual excess was treated as supporting evidence rather than a necessary condition; final delayed-obligation labels required local absence and semantic attribution in the diffs. The tree also mandated opening the GitHub commit page for all candidate delayed obligations. After independent coding, disagreements were resolved through consensus discussion. The critical label is debt attribution: it separates attributable delayed obligations from anchor-local hardening, background maintenance, pre-anchor hardening, classifier error, and unclear cases.

\subsection{Research Questions}

\textbf{RQ1.} How often do real changed-file diffs contradict commit-message warnings about missing anchors?

\textbf{RQ2.} Does post-anchor stabilization density exceed pre-anchor and non-anchor control density under broad, strict, and window-size sensitivity checks?

\textbf{RQ3.} What stabilization regimes explain dense anchor neighborhoods, and how often do candidate cases survive human validation as delayed obligations?

\section{RQ1: Anchors Are Common}

The first result is the reversal that motivated this study. The preliminary AnchorFinder scout sampled commit messages across repository phases and flagged repositories whose messages suggested feature growth without durable boundary terms. These warnings are mining shortcuts whose validity we test against changed-file diffs. Table~\ref{tab:phenomenon} summarizes the full-pool diff validation. Anchors appear in 95.0\% of repositories; stabilization follow-up appears in 99.7\%; and 94.7\% contain both. Commit-message missing-anchor warnings are contradicted in 81 of 84 checked cases.

\begin{table}[htbp]
\caption{Full-pool diff validation phenomenon check.}
\label{tab:phenomenon}
\centering
\begin{tabular}{lrr}
\toprule
Metric & Count & Rate \\
\midrule
Anchor observed in real diffs & 321 / 338 & 95.0\% \\
Stabilization follow-up observed & 337 / 338 & 99.7\% \\
Anchored stabilization observed & 320 / 338 & 94.7\% \\
Late-stabilization warnings confirmed & 254 / 263 & 96.6\% \\
Missing-anchor warnings contradicted & 81 / 84 & 96.4\% \\
Delayed-anchor warnings contradicted & 42 / 54 & 77.8\% \\
\bottomrule
\end{tabular}
\end{table}

Anchor prevalence is not an artifact of the sampling criteria. A skeptic could object that repositories were admitted partly on provider or agent-runtime evidence, and that two of the five anchor types (API/provider and AI-runtime) overlap with that evidence. We therefore audit prevalence across inclusion strata. Anchor presence is essentially flat whether or not a repository was admitted on provider/runtime evidence: 289 of 303 (95.4\%) provider/runtime repositories versus 32 of 35 (91.4\%) admitted without it, and 26 of the 28 repositories admitted only through AI-development or agent-documentation evidence still contain anchors. Most importantly, 266 of 338 repositories (78.7\%) contain a \emph{structural} anchor type (workflow/engine, storage/schema, or coordination-contract) that does not overlap with any inclusion criterion. Anchor commonality therefore reflects the projects, not the way the pool was assembled.

\textbf{Finding 1.} The value of this reversal is methodological, not merely a prevalence count: commit-message scouting overcalls missing anchors, and only changed-file diffs settle the question. Anchors are common, and common across sampling strata, so the stronger question is whether common anchors are stabilized well enough.

\section{RQ2: No Simple Post-Anchor Uplift}

We extracted 308 pipeline-selected candidate anchor events from the sampled-phase corpus and completed a contiguous-window validation. Table~\ref{tab:eventstudy} reports the 308-event run. The result is not a clean uplift. Broad stabilization file density is high in all windows. The median post/pre file-density ratio is 1.008, its bootstrap confidence interval includes 1.0 (1.0000--1.0224), and Cliff's delta for post versus pre is 0.0522.

\begin{figure*}[t]
\centering
\includegraphics[width=0.92\textwidth]{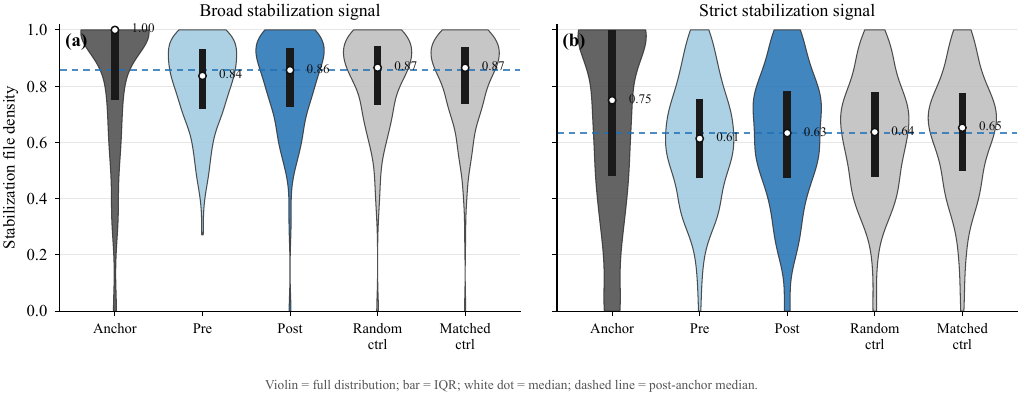}
\caption{Stabilization file-density distributions across the five window
types, under the broad (a) and strict (b) signals. The post-anchor
distribution (dark blue) sits among the pre-anchor and non-anchor control
distributions in both panels; its median (dashed line) does not exceed the
controls. The anchor commit itself is denser by construction because it is
the boundary-introducing change. $N=308$ for anchor/pre/post and $N=304$ for
each control.}
\label{fig:densities}
\end{figure*}

Figure~\ref{fig:densities} makes the overlap visible. Under the broad signal (panel a), the post-anchor median (0.86) falls between the pre-anchor median (0.84) and the two controls (0.87 each); the dashed post-anchor line cuts through all four neighboring distributions at the same level. The strict signal (panel b) shifts every window down by roughly 0.20 density points but preserves the same ordering and overlap. In neither panel does the post-anchor distribution separate from the controls.

\begin{table}[htbp]
\caption{Contiguous-window results for all 308 candidate anchor events.
File-density medians are shown for both the broad and strict stabilization
signals.}
\label{tab:eventstudy}
\centering
\tiny
\begin{tabular}{lrrrrr}
\toprule
Window & N & Med.\ commit & Broad file & Strict file & Mean file \\
\midrule
Anchor commit & 308 & 1.0000 & 1.0000 & 0.7500 & 0.8401 \\
Pre-anchor & 308 & 1.0000 & 0.8367 & 0.6138 & 0.8077 \\
Post-anchor & 308 & 1.0000 & 0.8571 & 0.6335 & 0.8144 \\
Random control & 304 & 1.0000 & 0.8661 & 0.6371 & 0.8226 \\
Matched control & 304 & 1.0000 & 0.8656 & 0.6524 & 0.8216 \\
\bottomrule
\end{tabular}
\end{table}

Paired within-event checks lead to the same conclusion (Table~\ref{tab:paired}). Post-anchor file density is greater than pre-anchor density in 155 of 308 events, lower in 137, and tied in 16; the median paired difference is 0.0020 with a bootstrap confidence interval crossing zero. When post-anchor windows are paired with random and matched non-anchor controls, the median difference is 0.0000 and post-anchor windows are not more often larger. These sign-test results are descriptive rather than causal, but they make the negative result easier to inspect.

\begin{table}[htbp]
\caption{Paired file-density checks for post-anchor windows.}
\label{tab:paired}
\centering
\tiny
\begin{tabular}{lrrrr}
\toprule
Comparison & Pairs & Post $>$ other & Med. diff. & Sign $p$ \\
\midrule
Post vs pre & 308 & 155 & 0.0020 & 0.3198 \\
Post vs random control & 304 & 133 & 0.0000 & 0.3718 \\
Post vs matched control & 304 & 138 & 0.0000 & 0.9522 \\
\bottomrule
\end{tabular}
\end{table}

The control-similarity audit reinforces this. Among 304 events with both controls, post-anchor density is greater than both controls in only 87 of 304 events (28.6\%). In 217 of 304 events (71.4\%), at least one non-anchor control is at least as dense as post-anchor; 183 events (60.2\%) are within 0.05 file-density points of a control. Thus post-anchor density alone is weak debt evidence.

An offline placebo-window audit gives a stronger specificity check. In 302 eligible events it builds 3260 cached nearby non-anchor pseudo-windows. Real post-anchor density exceeds the event's placebo median in 128 events (42.4\%) and all cached placebo windows in only 40 events (13.2\%); median actual-minus-placebo-median density is -0.0016. Thus shifted non-anchor windows often look just as stabilizing.

To avoid treating non-significance as absence, we also ran a practical effect-bound audit rather than a formal equivalence test. The 19 checks comprise four post/pre ratio confidence intervals (the primary run plus the 3-, 5-, and 10-commit windows), three paired median-difference intervals (post versus pre, random, and matched controls), and twelve Cliff's delta effect sizes across the same comparisons and window sizes; the bounds are declared a priori ($\pm5\%$ for the ratio, $\pm0.05$ file-density points for paired differences, and Cliff's 0.147 negligible threshold). All 19 audited intervals and effect-size summaries fall inside these small-effect bounds: ratio confidence intervals stay within [0.95, 1.05], with largest upper bound 1.0224; paired median-difference intervals stay within $\pm$0.05 file-density points, with largest absolute bound 0.0179; and the largest absolute Cliff's delta is 0.0522. Thus the RQ2 claim is not merely ``not significant''; under this automated metric the aggregate post-anchor uplift is practically small.

The strict sensitivity analysis tells the same directional story. The strict median post/pre file-density ratio is 1.0244, and 54.5\% of events are above 1.0. Strict densities are lower, as intended, but post-anchor windows remain close to pre-anchor and control windows; the strict-file column in Table~\ref{tab:eventstudy} places these medians beside their broad counterparts.

\textbf{Robustness checks.} The negative result is not an artifact of a single design choice. We re-ran the analysis across three window sizes, all five anchor types, and ten project-type strata.

Window size does not change the conclusion (Table~\ref{tab:windowsize}). Shrinking the window to 3 or 5 commits keeps the median post/pre ratio at 1.0, and even at 10 commits the median is 1.008 with a bootstrap interval that includes 1.0. The share of events with a post/pre ratio above 1.0 rises only gradually from 39.0\% to 51.6\% as the window grows---never approaching the dominance a real delayed wave would produce.

\begin{table}[htbp]
\caption{Window-size sensitivity of the post/pre file-density ratio.}
\label{tab:windowsize}
\centering
\begin{tabular}{rrrr}
\toprule
Window & Median ratio & Bootstrap CI & Ratio $>$ 1.0 \\
\midrule
3 commits & 1.0000 & [1.0000, 1.0000] & 39.0\% \\
5 commits & 1.0000 & [1.0000, 1.0087] & 45.8\% \\
10 commits & 1.0080 & [1.0000, 1.0224] & 51.6\% \\
\bottomrule
\end{tabular}
\end{table}

The conclusion also holds within each anchor type (Table~\ref{tab:anchortype}). Four of the five types have median post/pre ratios between 1.007 and 1.025---small and close to the aggregate. The one exception is informative in the opposite direction: coordination-contract anchors have a median ratio of 0.972, meaning post-anchor windows are \emph{less} dense than pre-anchor windows, and their dominant regime is background maintenance rather than saturation. Because anchor types are multi-label, the per-type event counts sum to more than 308. A leave-one-type-out check confirms that no single type drives the aggregate: removing any one type shifts the median ratio by at most 0.008.

\begin{table}[htbp]
\caption{Post/pre ratio by anchor type (multi-label; counts overlap).}
\label{tab:anchortype}
\centering
\scriptsize
\setlength{\tabcolsep}{4pt}
\begin{tabular}{lrrrl}
\toprule
Anchor type & $N$ & Med. ratio & Post/Pre med. & Dominant regime \\
\midrule
AI runtime & 158 & 1.013 & 0.86 / 0.84 & Saturation \\
API/provider & 147 & 1.020 & 0.86 / 0.83 & Saturation \\
Storage/schema & 96 & 1.025 & 0.81 / 0.80 & Saturation \\
Workflow/engine & 90 & 1.007 & 0.83 / 0.83 & Saturation \\
Coordination & 38 & 0.972 & 0.85 / 0.85 & Bg.\ maintenance \\
\bottomrule
\end{tabular}
\end{table}

Project type is equally inert. Across ten product domains, a leave-one-domain-out re-computation moves the median ratio by at most 0.0043, so no single domain, including the coding-agent/devtool stratum, explains the result.

\textbf{Finding 2.} The contiguous evidence does not support a naive ``anchor causes later stabilization increase'' claim. Stabilization work is dense before, during, and after candidate anchor events, and it is dense around non-anchor controls as well.

The full-run sensitivity result also clarifies what the broad signal is doing. The broad signal marks a median of 84--87\% of changed files as stabilization-related in the pre-, post-, and control windows alike---a near-saturation that is itself part of the measurement trap: when almost every file in every window is a plausible stabilizer, a post-anchor count cannot separate anchor-driven work from ordinary project motion. Crucially, this is not merely an artifact of a permissive classifier. The strict signal, which marks only a median of 61--65\% of files in the same windows, lowers every median but leaves the ordering and the overlap intact: post-anchor strict file density remains close to pre-anchor density and below non-anchor controls. The absence of a large post-anchor effect therefore survives both a recall-oriented signal that overcounts and a skeptical signal that discards generic ``fix'', broad API, and broad UI terms.

\section{RQ3: Stabilization Regimes}

\begin{figure}[t]
\centering
\footnotesize
\textbf{(a) Automated diagnostic regimes ($N=308$)}\\[1pt]
\begin{tikzpicture}
\begin{axis}[
  xbar, width=\columnwidth, height=3.15cm,
  enlarge y limits=0.16, xmin=0, xmax=162, bar width=5pt,
  symbolic y coords={Post-uplift,Anchor-local,Pre-anchor,Mixed,Background maint.,Saturation},
  ytick=data, nodes near coords, nodes near coords style={font=\scriptsize},
  axis x line=none, y axis line style={draw=none}, ytick style={draw=none},
  tick label style={font=\scriptsize}]
\addplot[fill=blue!55] coordinates {(136,Saturation) (84,Background maint.) (70,Mixed) (15,Pre-anchor) (2,Anchor-local) (1,Post-uplift)};
\end{axis}
\end{tikzpicture}\\[2pt]
\textbf{(b) Consensus debt attribution ($N=100$)}\\[1pt]
\begin{tikzpicture}
\begin{axis}[
  xbar, width=\columnwidth, height=2.9cm,
  enlarge y limits=0.18, xmin=0, xmax=54, bar width=5pt,
  symbolic y coords={Delayed oblig.,Pre-anchor,Background maint.,Anchor-local,Classifier error},
  ytick=data, nodes near coords, nodes near coords style={font=\scriptsize},
  axis x line=none, y axis line style={draw=none}, ytick style={draw=none},
  tick label style={font=\scriptsize}]
\addplot[fill=orange!75] coordinates {(45,Classifier error) (26,Anchor-local) (17,Background maint.) (9,Pre-anchor) (3,Delayed oblig.)};
\end{axis}
\end{tikzpicture}
\caption{Distributions behind RQ3. (a) The automated pipeline assigns each of
the 308 events to one of six diagnostic regimes; saturation and background
maintenance dominate. (b) Human consensus over the 100-case sample: 45 of 100
candidates are classifier errors and only 3 survive as attributable delayed
obligations. The automated regime distribution and the human attribution
distribution do not line up, which is the two-layer trap made visible.}
\label{fig:distributions}
\end{figure}
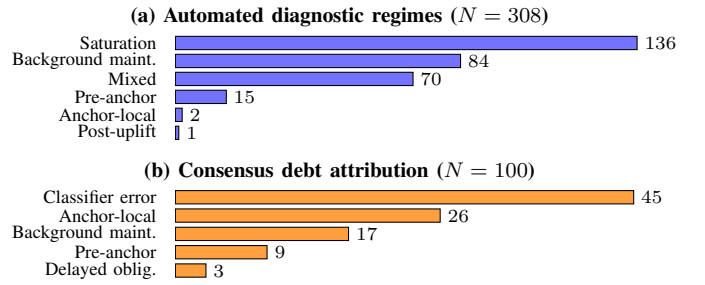

To make the mixed event-study result interpretable, we classify each event into diagnostic regimes using the broad window densities and controls. The 308-event population is dominated by two regimes: stabilization saturation and background maintenance (Figure~\ref{fig:distributions}(a)). Saturation means the anchor, pre/post windows, and controls are all dense. Background maintenance means the post-anchor window is dense but similar to controls.

Anchors sit in multi-category stabilization neighborhoods. AI-runtime and API/provider anchors are surrounded by CI/release, integration-provider, runtime-hardening, test, documentation, and UX work. No single category dominates the aggregate density.

To ground the regime taxonomy, two independent coders labeled a stratified 100-case subset drawn from the 308-event pool. Each coder received a blinded HTML evidence packet presenting the anchor commit, changed files, window densities, and commit URL---but hiding all automated diagnostic labels. The coders worked independently, following a decision tree that required inspecting the control comparison before any delayed-obligation label (with counterfactual excess as supporting evidence, not a gate) and opening the GitHub commit page for all candidate delayed-obligation and unclear cases. After independent coding, 35 disagreements on debt attribution were resolved through consensus discussion.

Inter-rater agreement before resolution was moderate for the critical field, debt attribution ($\kappa = 0.50$, raw 65\%), and lower for anchor validity ($\kappa = 0.39$, raw 60\%), consistent with debt attribution being the harder judgment.

Figure~\ref{fig:distributions}(b) shows the final consensus distribution. The largest category is classifier error, at 45 of 100 events: on close inspection the flagged commit is not a durable anchor. This mirrors the anchor-validity labels: only 32 of 100 commits are judged valid anchors, 23 weak, and 45 not anchors. The automated anchor signal is itself unreliable on this stratified sample. This does not contradict RQ1's finding that 321 of 338 repositories contain anchor evidence: a repository can contain genuine anchors while the automatically selected earliest candidate commit is an ordinary change that the pipeline misclassified. Anchor-local hardening explains another 26 events, background maintenance 17, and pre-anchor hardening 9. Only 3 events survive as attributable delayed obligations, identified by tracing the post-window stabilization back to the anchor boundary in the diffs; semantic attribution, not a density threshold, is what separates them.

The low survival count is the central consensus result: only 3 of the 100 stratified cases survived as attributable delayed obligations. Stabilization density is pervasive, but stabilization \emph{debt} in the temporal-attribution sense is rare. In the resolved queue, 97 of 100 cases received definitive non-debt explanations, showing how easily a post-only interpretation would overcall debt. All three surviving delayed obligations were contested during independent coding: in each case one coder labeled the event background maintenance while the other labeled it a delayed obligation, and the consensus discussion resolved toward delayed attribution after reviewing the diffs for semantic evidence. This confirms that delayed attribution is the hardest judgment, precisely the one that density cannot automate.

Crucially, the automated regimes do not locate the delayed obligations. Table~\ref{tab:regimexattr} cross-tabulates each event's automated regime against its consensus attribution. All three delayed obligations fall in events the pipeline labeled \emph{background maintenance} (the regime meant to signal ordinary maintenance), while \emph{post-anchor uplift}, the one diagnostic pattern a naive reading would treat as the debt candidate, contains a single event that the coders judged a classifier error. The automated diagnosis, like the density metric, does not isolate debt: the regimes are a descriptive routing layer, and only the semantic attribution step separates the three genuine obligations from the surrounding maintenance.

\begin{table}[htbp]
\caption{Automated regime versus consensus attribution (100-case sample).}
\label{tab:regimexattr}
\centering
\scriptsize
\setlength{\tabcolsep}{3.5pt}
\begin{tabular}{lrrrrrr}
\toprule
Automated regime & Clf.\ err & Anc.-loc & Backgr. & Pre-anc. & Delayed & $N$ \\
\midrule
Saturation & 17 & 12 & 0 & 1 & 0 & 30 \\
Background maint. & 8 & 3 & 13 & 3 & 3 & 30 \\
Mixed/unclear & 8 & 8 & 4 & 2 & 0 & 22 \\
Pre-anchor & 9 & 3 & 0 & 3 & 0 & 15 \\
Anchor-local & 2 & 0 & 0 & 0 & 0 & 2 \\
Post-anchor uplift & 1 & 0 & 0 & 0 & 0 & 1 \\
\bottomrule
\end{tabular}
\end{table}

A natural concern is whether the no-uplift finding (RQ2) is an artifact of invalid anchors diluting the signal. Table~\ref{tab:anchorsens} tests this by stratifying the 100-case sample by consensus anchor validity.

\begin{table}[htbp]
\caption{RQ2 sensitivity to anchor validity (100-case validation sample).}
\label{tab:anchorsens}
\centering
\small
\begin{tabular}{lrrrr}
\toprule
Anchor validity & $N$ & Med.\ post/pre & Post $>$ pre & Delayed \\
\midrule
Valid anchor & 32 & 1.029 & 17/32 & 3 \\
Valid + weak & 55 & 0.998 & 24/55 & 3 \\
Not anchor & 45 & 0.995 & 20/45 & 0 \\
All 100 & 100 & 0.997 & 44/100 & 3 \\
\bottomrule
\end{tabular}
\end{table}

Even among the 32 events judged valid anchors, the median post/pre ratio is 1.029, essentially no uplift, and only 8 of 32 post-anchor windows exceed both controls. All three delayed obligations fall in this group, confirming that they are genuine anchor events, but they do not generate a detectable aggregate signal. The no-uplift conclusion is not an artifact of classifier error: it holds in the valid-only, valid-plus-weak, and not-anchor subsets alike.

The regimes are diagnostic failure-mode labels rather than debt predictors. In a saturation regime, a post-anchor window can look risky even when the entire local development neighborhood is already stabilization-heavy. In a background-maintenance regime, a post-anchor window can contain many tests, docs, or CI edits while comparable non-anchor windows show the same pattern. In a pre-anchor regime, the project may have hardened before the boundary was made explicit, so post-anchor debt would be the wrong temporal interpretation. In an anchor-local regime, the stabilizers appear synchronously with the anchor and should be treated as evidence of good engineering practice rather than delayed obligation. The mixed regime is the one that should receive the most human review: it is where density differences, changed-file semantics, and commit ordering are not enough to settle attribution.

This taxonomy is the paper's main technical step between measurement and action. It transforms the question from ``did stabilization appear after the anchor?'' to ``which counterfactual would make this post-anchor work debt?'' For saturation, the counterfactual asks whether the anchor created a new burden relative to an already-hardening neighborhood. For background maintenance, it asks whether the same work would have happened without the anchor. For pre-anchor hardening, it asks whether the anchor is actually the result of prior stabilization. For anchor-local hardening, it asks whether the expected stabilizers were delivered in the same change set. Only after answering those questions can a reviewer decide whether a later test, workflow, guard, or documentation change is a delayed obligation rather than ordinary project motion.

Three resolved cases illustrate how regime attribution works in practice.

\textbf{Case A: Background maintenance (PAE-013, vibeframe).} The anchor adds a CLI package for headless video editing, introducing API/provider, coordination, and workflow boundaries. Post-anchor file density is 0.82, but random-control density is 0.89 and matched-control density is 0.91. A post-only metric would flag this event: seven stabilization categories appear in the post window. Yet the controls show that non-anchor windows are equally dense. Both coders and the consensus labeled the case background maintenance: the density pattern (post below both controls) rules out a density-based delayed-obligation claim, and the consensus found no semantic evidence to override that pattern.

\textbf{Case B: Classifier error (PAE-034, llm-for-zotero).} The automated pipeline flagged this commit as an AI-runtime anchor, and post-anchor density reaches 1.0, a pattern a post-only metric would read as strong debt evidence. But the commit is a bug fix (``Malformed Tool Argument and Trace Redaction Fix''), not a durable boundary: its 13 changed files (five test files and eight source files) repair existing tool-argument handling and trace rendering rather than introducing a contract that future work must respect. Both coders and the consensus labeled it not an anchor, the single most common outcome (45 of 100), and a reminder that the automated anchor signal itself must be filtered by hand before a debt question is even meaningful.

\textbf{Case C: Delayed obligation (PAE-120, vector-graph-rag).} The anchor is a provider/runtime refactor that replaces the project's local document model with \texttt{langchain\_core}'s \texttt{Document} and swapping the LangChain OpenAI wrapper for the native OpenAI API across the extractor, reranker, and RAG interfaces. Density is inconclusive: post-anchor file density (0.74) sits between the random control (0.72) and the \emph{denser} matched control (0.95), so a control comparison would not flag the event. What identifies it as a delayed obligation is semantic: the interface change landed without its full support, and the post-window commits add the integration tests, dependency pins, configuration, and runtime guards that the new provider boundary requires, work the consensus tied directly to the refactor by reading the diffs. This is one of only 3 surviving cases, and, unlike Cases A and B, the judgment could not be made from densities alone.

\textbf{Finding 3.} The phenomenon is stabilization regimes, not a simple post-anchor spike. Human validation confirms that only 3 of the 100 stratified events are attributable delayed obligations; the remaining 97 have definitive non-debt explanations.

\section{Discussion}
\label{sec:discussion}

\subsection{The Measurement Trap}

The central finding is a two-layer measurement trap. The first layer is \emph{anchor extraction}: the automated pipeline mistakes ordinary bug fixes and refactors for durable boundary events. In the 100-case validation sample, 45 candidate anchors are not anchors at all, and only 32 are judged valid by human coders. The second layer is \emph{density attribution}: even among valid anchors, post-anchor stabilization density does not distinguish delayed obligations from routine maintenance. Valid anchors have a median post/pre ratio of 1.029 (Table~\ref{tab:anchorsens}), essentially no uplift, because high-visibility repositories maintain dense stabilization work throughout their commit history, not only after anchors.

Both layers are needed to understand the trap. If the first layer were the only problem, filtering to valid anchors would recover a usable density signal. It does not: post-anchor density remains indistinguishable from controls even in the valid-only subset. Conversely, if the second layer were the only problem, one might still trust automated anchor extraction as a reliable starting point. It is not: 45 of 100 candidates fail human inspection. A naive post-density threshold at 0.70 would flag 243 of 308 events (78.9\%), but human validation shows that only 3 of 100 stratified cases are delayed obligations.

The aggregate null does not mean that stabilization debt never occurs. Across the full 308-event population the post-anchor signal is statistically indistinguishable from pre-anchor and control windows, which is consistent with attributable debt being rare and heterogeneous rather than absent: a few genuine cases diluted among pervasive maintenance. That combination defeats an aggregate density metric. The signal exists, but it is too sparse and unevenly distributed to survive averaging, so a per-event density threshold reports mostly false positives. The stratified human queue calibrates this interpretation (it shows which failure modes dominate) but is not itself a prevalence estimate.

This is not a null finding in the ``we looked and found nothing'' sense; it falsifies a specific measurement approach. Any repository-mining study that reports post-anchor density as evidence of technical debt risks the same conflation, which the regime taxonomy and human-calibrated consensus labels are meant to avoid.

\subsection{From Measurement Error to Scope Drift}

The overcall is not only a statistical error but an action error: a coding agent or AI reviewer that treats dense post-anchor work as deferred obligation may recommend unnecessary repairs, test additions, or refactors, expanding a bounded task beyond what the developer requested. This path from false measurement to false action is a mechanism for agent-induced scope drift.

Our repository corpus does not contain user-agent interaction transcripts, so it cannot directly observe conversation-level drift. However, it identifies the code-level endpoint where such drift becomes measurable: new anchors surrounded by dense stabilization regimes. The regime taxonomy provides a structured routing step that a naive density metric omits. The decisive test for opening a delayed-obligation ledger item is \emph{semantic attribution}: later stabilizers that can be tied to the specific anchor boundary, supported by \emph{local absence} (the expected stabilizers were missing or incomplete when the anchor landed). Density comparison is only a weak filter. \emph{Counterfactual excess}, post-anchor stabilization exceeding pre-anchor and non-anchor controls, is neither necessary (two of our three genuine delayed obligations show no such excess) nor sufficient (most windows that do show it are ordinary maintenance). Absent a semantic tie, a case should be labeled saturation, background maintenance, pre-anchor hardening, anchor-local hardening, or unclear, not debt and not an occasion for corrective action.

\subsection{Implications for Repository Mining}

The design is intentionally within-pool: we study what happens inside high-visibility AI-evidence repositories, not whether they differ from non-AI ones. A comparative claim would require a matched baseline, and our feasibility probes show that neutral 2026 GitHub queries are heavily contaminated by AI surfaces, making clean matching non-trivial. The trap is unlikely to be AI-specific: its mechanism (pervasive stabilization across an active project's whole commit history) should hold for any such repository, and we study it in AI-evidence projects only because that is where post-anchor debt metrics are now being applied. Future comparative studies should apply this protocol only after regime-aware attribution is validated, to avoid comparing two unreliable classifiers.

\subsection{Practical Implications}

The regime taxonomy suggests a practical anchor-ledger workflow for code reviewers and AI coding agents. When a boundary-changing commit appears, the reviewer asks: which stabilizers are expected for this anchor type, and which are present in the local window? A ledger entry is opened when an anchor is introduced, updated as stabilizers appear, and closed only after a reviewer can explain whether the surrounding work is synchronous hardening, background maintenance, or a delayed obligation.
\begin{table}[htbp]
\caption{Ledger actions generated from anchor types.}
\label{tab:ledgeractions}
\centering
\scriptsize
\setlength{\tabcolsep}{3pt}
\begin{tabular}{p{0.24\columnwidth}p{0.33\columnwidth}p{0.33\columnwidth}}
\toprule
Anchor type & Expected stabilizer & Ledger action \\
\midrule
API/provider & Contract tests, fallback guards & Open provider-risk entry \\
AI/runtime & Model config, timeout handling & Check runtime failure modes \\
Workflow/engine & CI jobs, queue fixtures & Require executable gate \\
Storage/schema & Migration and rollback tests & Track data-safety entry \\
Coordination & Agent docs, release checks & Verify contract enforcement \\
\bottomrule
\end{tabular}
\end{table}

Table~\ref{tab:ledgeractions} is operational rather than predictive: it asks a review question without claiming that every missing stabilizer is debt.

\section{Threats to Validity}

\textbf{AI-evidence is not AI-authorship.} The repositories contain explicit AI-related evidence, but we cannot claim that all changes were generated by AI. The study is about AI-evidence repository evolution, not verified AI-authored commit quality.

\textbf{GitHub sampling bias.} GitHub search favors visible public projects, and our star/commit thresholds exclude smaller repositories. We report this as scope, not universal behavior.

\textbf{Heuristic classification.} Broad signals intentionally favor recall and can overcount ordinary maintenance. The strict sensitivity analysis reduces that risk but does not replace schema-bound semantic judgment.

\textbf{Attribution difficulty.} Not every later stabilizer is debt. The final claim depends on separating generic maintenance from delayed obligations tied to an anchor.

\textbf{Single-event abstraction.} Each repository contributes one earliest qualifying candidate anchor event, which makes the study reproducible and avoids author-selected salient events but does not model later anchors. Selecting the earliest anchor may also bias the sample toward greenfield boundaries introduced together with their stabilizers, inflating the share of anchor-local hardening relative to later retrofit anchors. The results therefore concern selected anchor neighborhoods, not each project's complete anchor history.

\textbf{Human validation scale.} The 100-case sample is stratified to cover all six regimes but is not a random population sample; we report it as a consensus calibration, not a prevalence estimate. Both coders have software-engineering backgrounds, and coders from other domains might draw the anchor-validity boundary differently.

\section{Data and Artifact Availability}

The supplementary material is organized in three layers. The \emph{sampling layer} records the 338-project accepted pool, the 357 rejected candidates, and provenance metadata. The \emph{measurement layer} records the 308 contiguous event windows, broad and strict sensitivity outputs, the control, placebo, and practical-bound audits, the robustness checks, and the diagnostic-regime labels, together with the raw per-commit diff cache for all 308 events. The \emph{validation layer} includes the blinded coding protocol and packet generator, the two independent coder codings, the consensus labels, and the 35 resolved disagreements. Third-party repository identifiers and commit metadata are retained for reproducibility, and the package carries no author-identifying information. The cached tables and data reproduce the paper's empirical tables and Figures~\ref{fig:densities} and~\ref{fig:distributions} without re-querying GitHub.
\noindent All supplementary materials are available at \texttt{https://doi.org/10.5281/zenodo.21099289}.
\section{Conclusion}

This study began with a worry that AI-evidence repositories might lack durable architecture anchors. Real diffs show the opposite: 321 of 338 repositories contain anchor evidence. A second intuitive story, that stabilization spikes after anchors, also fails: a controlled 308-event experiment finds no post-anchor uplift.

These reversals reveal a two-layer measurement trap. Automated anchor extraction misclassifies many candidate events, and even among valid anchors in the calibration sample, post-anchor density does not distinguish debt from routine maintenance. Six stabilization regimes explain why the metric fails, and human validation ($\kappa = 0.50$) calibrates them: only 3 of 100 stratified events survive as attributable delayed obligations.

Repository-mining studies should therefore not report post-anchor density as evidence of technical debt without controlled designs and regime-aware attribution. The risk extends beyond measurement: an agent or reviewer acting on false debt labels may recommend unnecessary repairs or scope-expanding changes, turning a measurement overcall into an engineering mistake.

\enlargethispage{2\baselineskip}
\bibliographystyle{IEEEtran}
\bibliography{references}

\end{document}